Chapter 8

# Model Driven Testing of Time Sensitive Distributed Systems

**Abstract** In this paper we demonstrate an approach to model structure and behavior of distributed systems, to map those models to a lightweight execution engine by using a functional programming language and to systematically define and execute tests for these models within the same technology. This is a prerequisite for a smooth integration of model based development into an agile method. The novelty of this paper is the demonstration, how composition and state machine models for distributed asynchronously communicating systems can easily be mapped to a lazy functional language and then using standard testing techniques to define test on those programs. In particular distributed timing aspects and underspecification can be treated accordingly within a functional language, using a certain style of functions.

## 8.1. Model Driven Testing

Software engineering in recent years has come up with a larger portfolio of techniques and methods to measure and improve or ensure quality of software products. Among those methods we have available inspection and review techniques for all artifacts that are produced during development. These techniques require clear qual-

---

Chapter written by Borislav Gajanovic, Hans Grönniger and Bernhard Rumpe

ity criteria as well as an appropriate process that clarifies the order of artifacts to be delivered and examined.

Among the most promising techniques today, however, is the execution of "running" versions of system descriptions and the check of the execution result against the desired result. Agile methods [BEC 99a, COC 02, BEC 01, RUM 04] have furthermore successfully demonstrated that a full automation of the testing process is of high value in any larger and quality driven project. Only automated tests can efficiently be reused by developers.

The portfolio of testing techniques has become large. It ranges from functional tests on abstract specifications over black box tests, stress tests, random tests down to glass box tests derived from the code. As testing has become a powerful technique, it was a natural idea to lift the use of testing techniques from the late coding phase to earlier phases, where fixing errors is less costly. However, starting early with testing means, we need executable artifacts early in the development process. If requirement specifications or at least design and architectural artifacts are executable, they need to be defined (a) precisely and (b) describe not only structural aspects but also behavioral issues. This calls for precisely defined modeling languages and for tools dealing with those languages. Tools need to be able to analyze well-formedness (thus checking context conditions) and to animate the model or to be able to map the model to code.

The necessary properties of a model strongly depend on the context of the model usage. Models can be used for constructive or for test code generation. The desired properties here are fundamentally different. In case of a constructive code generation, we do have a compiler and as "side effect" modeling is equal to implementation. We can reuse the generated code as implementation if it is not only compatible with the development computers, but also with the target and is efficient enough. Instead, if we want to generate testing code, we do neither need complete and therefore very detailed models, nor do we need to restrict us to an executable modeling language. To understand the difference, consider a post condition for a method of the form $a^n = b^n + c^n \land a, b, c, n > 2$. Conditions like that are very easy to be checked, but it is usually extremely hard to construct a program that finds such values. So the choice of appropriate modeling languages and styles during the various development stages is important.

Statecharts [HAR 87] are among the most interesting forms of descriptions for executable behavior. The use of underspecified Statecharts [PAE 94, KLE 97] with transition conditions and actions in (almost) first-order-logic allows to generate checking code only, but not constructive code. A clever choice of appropriate modeling techniques and their underlying semantics is therefore inevitable.

For early checking of requirements, it is necessary to have a concise modeling technique and an efficient way of simulating the models at hand. In this paper we demonstrate an approach used to early validate requirements on distributed systems. This approach is exemplary, but demonstrates what can be achieved when using

- concise, compact modeling techniques in early phases,
- tools for transforming modeling techniques to executable languages, and
- a lean process to develop and use the models during the development adequately.

To demonstrate the approach, we choose a relatively simple, but not too simplistic protocol, namely the Alternating Bit Protocol (ABP) [BRO 01]. The two modeling languages used for structure and behavior are adapted versions of UML diagrams: the composite structure diagrams and (flat) Statecharts respectively state machines. As the protocol is useful for distributed asynchronously communication systems, we choose the use of streams as underlying technical domain, because it offers a very precise semantics as well as a good integration of composition, refinement and various styles of specification [BRO 01, RUM 97]. For a lean development, we use a Haskell [THO 99, BIR 98] interpreter to simulate the models. Haskell e.g. provides lazy evaluation of lists, which allows to almost perfectly simulate streams as potentially infinite observations over communication channels.

The remainder of this paper introduces the concept of distributed systems (Section 8.2), the Alternating Bit Protocol (Section 8.3), the general approach on testing (Section 8.4) and the application of testing strategies on the ABP (Section 8.5). Finally our findings are discussed (Section 8.6).

## 8.2 Asynchronous Communication in Distributed Systems

In this section we briefly introduce the used semantic framework that describes on asynchronous communication as underlying communication principle of our framework. In a distributed system, we do have active components that communicate with each other through asynchronous sending and receiving of messages. We assume communication is based on unidirectional channels. For a precise modeling of their behavior, we use observation histories that describe what happens on these channels over time using the mathematical concept of streams. Streams are thus used to model the message-flow over those channels. The behavior of a component is specified through a description of the input-output relation on these streams.

A stream is a finite or an infinite sequence of messages from some fixed finite set of possible messages (type). A stream-based specification of a component consists of a black box and an arbitrary (but finite) number of directed input and output channels. In figure 8.1 a composition of a system from several components is shown. With a single stream we describe the history on one channel, a component behavior is given by a function mapping its input streams to output streams. Certain restrictions apply on these functions to ensure the behavior is well-defined. E.g. a component (and thus a function) cannot undo messages that have been emitted, and it cannot react on future input (and thus predict the future).

For any realizable component there is a stream-processing function with the same input/output behavior. Any state machine (with possibly infinite number of states) whose state transition function acts on messages appearing at the input and is yielding appropriate streams on the output can also be defined via a stream-processing function [RUM 97, RUM 99]. Together with the fact that semantic definition through stream processing functions support composition of components via channels, we're able to use state machines and a variant of composite structure diagrams as a comprehensive specification language on a well-defined semantic basis. These stream processing functions are the primary concept to be mapped to Haskell for execution and testing. Furthermore, specification languages are particularly powerful if they allow us to abstract from implementation details and provide concepts for underspecification (alternatives, etc.). In terms of our underlying formalism, this means that a specification does not correspond to a single function, but to a set of functions that describes a set of possible implementations.

The list given below describes a basic set of operators on streams useful for the specification of the components. For a comprehensive introduction to the stream-based specification and development technique see [BRO 01].

- `[M]` - The set of all streams over a set of messages *M*.
- `[]` - The symbol for the empty stream.
- `[c]` - The stream consisting of a message `c`.
- `head s` - Yields the first message of a non-empty stream *s*.
- `tail s` - Returns the rest of a nonempty stream.
- `#s` - Returns the length of a stream `s` (may be $\mathbb{N} \cup \{\infty\}$).
- $s_1$ `++` $s_2$ - The concatenation of $s_1$ and $s_2$.
- `filter S s` - Filtering a stream `s` with respect to the members of a set S.

Please note the special case `#`$s$ = $\infty$ $\Rightarrow$ $s_1$ `++` $s_2$ = `s1`. Also note that some operators like length `#` count infinite things and therefore cannot be used in an implementation.

As an additional concept, we need the idea of time to model time-sensitive systems. Time can easily be modeled through the introduction of a special "message" - here symbolized with `Tk` and pronounced as tick - into the underlying set of messages. For that purpose we define an operator `T` for the introduction of so-called timed streams over a set of messages *M*:

$$T\ M = \{\ s \mid s \in [M \cup \{Tk\}]\ \wedge\ \#(\texttt{filter}\ \{Tk\}\ s) = \infty$$

A tick in a stream represents the incrementation of a global digital clock in the system. The time is never ending so there are infinitely many ticks in a timed stream. Timed streams are nothing more than normal infinite streams with a special structure. Hence, all operations introduced in the previous section can be applied to the timed streams as well.

### 8.3 The Alternating Bit Protocol

In this section we describe how to apply our approach to a simple version of the Alternating Bit Protocol that can be found in [BRO 01, BRO 93]. We use the Alternating Bit Protocol here as a simple example of a time sensitive distributed system which transmits data safely over unreliable media. The black box specification of the ABP is simple: Abstracting from possible delays, the ABP is the identity. An appropriate specification of the ABP is therefore given through a set of stream processing functions:

$$\texttt{ABP} \subseteq \texttt{T}\,M \to \texttt{T}\,M$$

where for any input `inp` $\in$ `T M` we have the abstraction of timing information on input and resulting output is identical:

$$\texttt{filter}\,M\,(\texttt{ABP}\,\mathit{inp}) = \texttt{filter}\,M\,\mathit{inp}$$

#### *8.3.1 Informal Description of the ABP Components*

The problem to be solved by the ABP is to transmit information over an unreliable medium. Thus the ABP must be decomposed (in a simplified version) like shown in figure 8.1. The system consists of a sender, a receiver and two transportation media. Both versions of the media are identical, except for the transported kind of messages. The medium is assumed to be given (e.g. in form of the internet or a bus). This means the model of the medium describes a given situation. In contrast, the sender and the receiver have to be defined in such a way, that the overall specification is correct. Our task is therefore to model all four components accordingly.

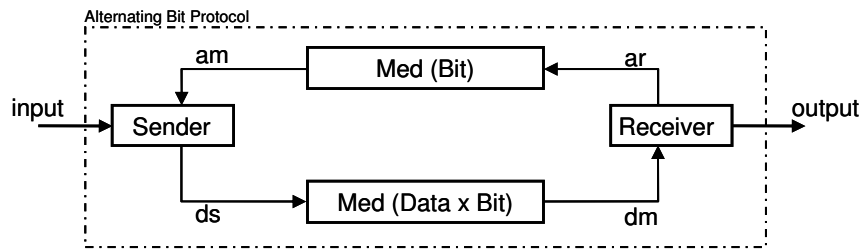

**Figure 8.1.** *The alternating bit protocol as a composition of its components*

The medium on the bottom of the figure is used to transport signed data items from the sender to the receiver, whereas the upper medium transports the acknowledgements back to the sender. The media however occasionally lose or delay messages (see details below). To detect loss of messages, the basic idea is to tag the message by a number and let the receiver acknowledge the tagged message by replying the number. If a message or acknowledgement gets lost, the sender repeats the message after a while (timeout).

We assume there is only one message in transmission: the sender stores forthcoming messages in a buffer until the last sent one is acknowledged. For reasons of efficiency, the message numbering can now be replaced by a single bit that alternates for each message. Thus each data item from the input channel is alternatingly signed by the sender using a bit. The receiver returns the bit and writes the corresponding data to the output. If a message or an acknowledgement was delayed too long, then the sender resends the message. Details like what happens when the delay is too long can be deduced from the below given specification. To model an unreliable, but not demonic transportation medium, we assume the medium to have the following properties:

1. If a message is sent infinitely often, then it will pass infinitely often.
2. The medium does not change the order in which the messages have been sent.
3. The medium does not duplicate messages or alters the message content.

Number 1) is a typical fairness condition on a medium. In other words it is always possible to transmit an item through an unreliable medium within some finite (but unknown) number of transmission attempts. In the following section we give a formal, stream-based specification of the above system.

*8.3.2. Stream-Based Specification*

The structure of the complete system is already given through the composition of its components in figure 8.1. This diagram is a variant of UML's composite structure diagrams that allows us to specify a complete system by the composition of its components (resp. "parts"), the components communicate asynchronously via unidirectional channels as described in section 8.2. In the reminder of this section we can therefore concentrate on the specification of each individual component. We use state machines to model the component's behavior. Since we allow a potentially infinite number of states, we graphically partition the state space into equivalence classes. These classes are given as invariants over variables inside the (graphically visible) states. To precisely represent the state space, we define the data types of these variables in form of a box (similar to a class definition in class diagrams). Transitions are of the form `{pre} i /o1,..,on {post}` where `pre` and `post` denote the transition pre- and post-condition, `i` represents the input and `o1,..,on` is a sequence of outputs.

**Sender:** The sender is a time sensitive, "intelligent" component of the protocol. In fact it is the most complicated component in the system. We specify the main functionality of the sender given above through the following state machine.

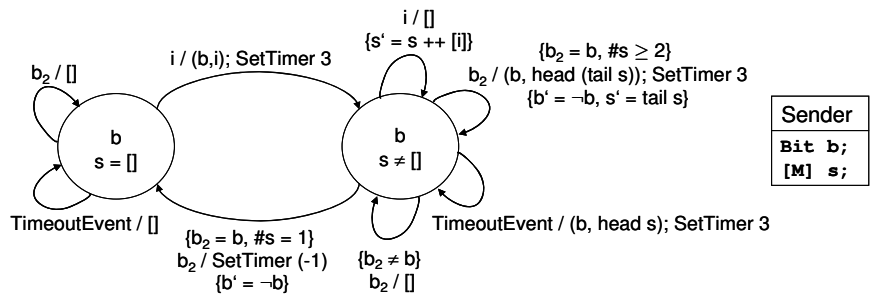

**Figure 8.2** *State machine for the sender component*

The sender component owns two input channels: the data input channel (*input*) and the acknowledgement channel (*am*). For reasons of readability, we do not add the channel names to the inputs. Although we do have two input channels, we can distinguish inputs from the channels, as their types are disjoint (Boolean vs. abstract type Message). For an understanding of the specified behavior, the reader may assume, we conceptually "merge" the two input streams to be able to let the state machine fire on the union of incoming events from both channels. The senders state space basically contains the unbounded buffer for data items still to be sent, and by the expected acknowledgement bit. The handling of time is assumed to be done by a timer, that "controls" the state machine by filtering `SetTimer`-events from its output and injecting `TimeoutEvent` to its input. So formally, input and output of the state machine (but not the sender itself) are extended by `TimeoutEvent` resp. `SetTimer`-messages.

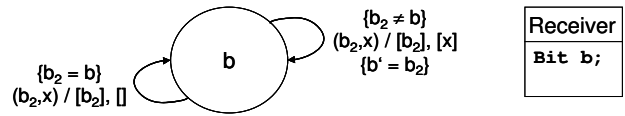

**Figure 8.3** *State machine for the medium component*

**Medium:** For specification of unreliable media in the described form either set based specifications or oracles are used. The unreliability of the medium is a given property which cannot be overcome, but needs to be modeled. Therefore we handle oracles for both media as two inherent system parameters. An oracle is an infinite binary stream which predicts the behavior of a particular medium over a complete communication history. As the media are time insensitive components, there is no need to consider time in the corresponding specifications. Instead, this is done systematically (and then automatically when mapped to code) through a simple time extension of the transition-function of a corresponding state machine. This extension just ignores incoming time events. The following state machine specifies untimed polymorphic media with the above properties. However, an additional predicate `#(filter {1} o) = ∞` is needed to describe the desired fairness property as an assumption on any oracle `o`.

**Receiver:** Just like the media the receiver also is a time insensitive component and can straightforwardly be specified as shown in the following figure.

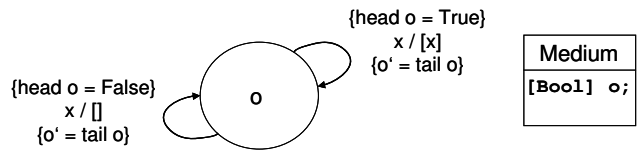

**Figure 8.4 State machine for the receiver component**

### *8.3.3. A Mapping To Haskell*

To be able to run tests based on the above given specification, we need an compiler or interpreter that resembles the underlying communication primitives appropriately. We choose the functional language Haskell, for a number of reasons. Most impor-tant, the Haskell type of lists can be directly taken to represent streams. Therefore, we now show how to automatically transform the system specified above into an executable Haskell [THO 99, BIR 98, JON 03, HUD 99] program. First we need appropriate data types. We introduce the type `Bit` to represent the corresponding type in the specification. Then the type constructor `Ticked` is introduced to inject the time in streams as described in the previous section. A timed stream of as is abbreviated with `T a`. Oracles are represented as streams of Boolean values. The type of a state machine's delta function is aliased with `DeltaFct s i o` where `s` is the state, `i` the input and `o` the output type. As the concept of union of sets is not directly present in Haskell, we use the type constructor `MergeAB a b` to represent the union of two types `a` and `b`. The rest of data type definitions allow the addition of a timeout event or the setting of a timer value to a given type `a`.

```
type Bit = Bool
data Ticked a = V a | Tk
type T a = [Ticked a]
type Oracle = [Bool]
type DeltaFct s i o = (s -> i -> (s,[o]))
data MergeAB a b  = A a | B b
data TimerOut a   = MsgO a | SetTimer Int
data TimerIn a    = MsgI a | TimeoutEvent
```

**Runtime system:** To keep the generated "code" as simple as possible, to facilitate reuse and to avoid generation of the same functions over and over again, the code is generated against a runtime system. This is quite common, as usually some concepts of the source domain need to be simulated in the target domain. Due to space constraints, only the type definitions of functions we use are shown here. First, we use a function to execute state machines. `execSTM` takes a state, a delta function and a list of inputs and produces a list of corresponding outputs. `execSTM` works on timed transition functions as well as untimed ones. To inject time (for time-insensitive descriptions) the `timedDelta` function is used. It basically adds behavior that resembles a loop transition to every state, consuming a Tick at the input and emitting a Tick at the output. The `mergeAB` function merges two streams, thus covering the merge of messages from several inputs, while retaining the source channel in the constructors name (see merge type above). The sender contains transitions that control a timer. In fact, we're able to use timers in a state machine. The `addTimer` function takes a delta function containing these timer control statements and generates a timed delta function that reacts to the corresponding events. Please recall that the timed function using ticks to model time is the final goal.

```
execSTM    :: s -> DeltaFct s i o -> ([i] -> [o])
```

```
timedDelta :: DeltaFct s i o -> DeltaFct s (Ticked i) (Ticked o)
mergeAB    :: T a -> T b -> T (MergeAB a b)
addTimer   :: DeltaFct s (TimerIn i) (TimerOut o) ->
                          DeltaFct (s,Int) (Ticked i) (Ticked o)
```

**Sender:** We use the sender component to demonstrate the mapping of a state machine to Haskell that interacts with our runtime system, because it has an untimed specifications, but uses a timer to check timeouts.

The `SenderState` consists of the last sent bit and the unbounded message buffer. The input to the sender `SenderIn` is either a message of type `a`, the acknowledgement bit or some timeout event (generated by the internal timer). `SenderOut` defines the sender output as an alternating bit together with a message of type `a` but may also be a setting of a timer. The function `senderDelta` now basically is a one-to-one mapping from the specification (see Figure 8.2). We use pattern matching on the current state and input to define the delta function. Note that transitions with same inputs and states but with different pre-conditions are handled in an if-then-else statement. Timer functionality is added to the sender's delta function using `addTimer`, giving the new delta function `sDelta'`.

According to figure 8.1 the `sender` takes an input stream and a stream of acknowledgements `am` and emits a stream of messages with an alternating bit. We have chosen to merge the input and the bit stream (in accordance with the specification). To complete the mapping, the sender state machine is executed on this merged stream, starting with the alternating bit set to `True`, an empty buffer and disabled timer ($-1$).

```
type SenderState a = (Bit,[a])
type SenderIn    a = TimerIn (MergeAB a Bit)
type SenderOut   a = [TimerOut (Bit,a)]

senderDelta:: SenderState a -> SenderIn a ->(SenderState a,SenderOut a)
senderDelta (b,   []) (MsgI (A i)) = ((b,[i]),[MsgO(b,i), SetTimer 3])
senderDelta (b,   xs) (MsgI (A i)) = ((b,xs++[i]),[])
senderDelta (b,   []) (MsgI (B b2)) = ((b,[]),[])
senderDelta (b, x:xs) (MsgI (B b2)) =
      if      b/=b2   then ((b,x:xs),    [])
      else if null xs then ((not b,[] ), [SetTimer (-1)])
      else ((not b,xs),   [MsgO (not b,head(xs)),SetTimer 3])
senderDelta (b,[])   TimeoutEvent = ((b,[]),[])
senderDelta (b,x:xs) TimeoutEvent = ((b,x:xs), [MsgO (b,x),SetTimer 3])
sDelta' = addTimer senderDelta
```

```
sender :: T a -> T Bit -> T (Bit,a)
sender is am = (execSTM ((True,[]),-1) sDelta') (mergeAB is am)
```

For the **Medium** and **Receiver** component only the type definitions are shown. They also correspond to the respective state machine specification and therefore illustrate that the resulting system will be composed. Please note, that the receiver component has to start with the same acknowledgement bit `True` as the sender (see above).

```
mediumDelta :: Oracle -> a -> (Oracle,[a])
medium :: Oracle -> T a -> T a

receiverDelta :: Bit -> (Bit,a) -> (Bit, [Bit],[a] )
receiver :: T (Bit,a) -> (T Bit,T a)
```

*8.3.4 Executing the Model*

Finally the complete system is just a composition of its components (compare fig. 8.1). Since we have a feedback loop in the system, we need to insert a start-up delay. This delay is inserted as an extra Tick prepended to the second medium's output.

```
abp :: (Oracle,Oracle) -> T a -> T a
abp (os1,os2) is = out
   where (as,out) = receiver dm
         dm       = medium os1 ds
         am       = Tk:(medium os2 as)  -- delay for feedback
         ds       = sender is am
```

**8.4. Strategies for Testing Distributed, Asynchronously Communicating Systems**

Based on the model of our system and our mapping of the model to Haskell we can now discuss strategies for testing distributed, asynchronously communicating systems. After some general remarks on testing, we discuss how to systematically derive test cases from our component models as well as from the composition model. Then we demonstrate how these test cases can be implemented in (or generated for) Haskell in a lightweight manner, i.e. not depending on any complex third-party test framework, but exploiting the special features of higher-order functional programming languages. First, we recall some general rules for developers and testers [MYE 79] that also apply for testing of models.

1. The tests should be regression-enabled. That means that if the program changes, the test can be replayed easily to check if all tested properties still hold. How this is supported by a lightweight test infrastructure is mainly described in section 8.5.

2. The tests should be kept local. This is a general problem of complex software where it is often the case that testing some method of an object means also executing many other interlocked methods of other objects with possible side effects. Object oriented systems do provide their own solutions through substituting parts of the context through stubs.

    Keeping test cases focused also applies to functional programs, where this kind of substitution through subclassing unfortunately does not exist. In the context of Haskell, we examine techniques that help programmers to write effective test cases for the generated functions.

3. The quality of the test cases should be assessed. A wide range of techniques exists to assess the quality of the tests. Since we use state machines for component specification, well known coverage criteria like transition coverage etc. can be applied directly [BIN 99].

### *8.4.1. Rules for Testing of Distributed Functionally Specified Models*

Testing models for distributed, asynchronously communicating programs implemented in Haskell as we did, in principle means testing functions.

In general we can benefit from the fact that these functions are side-effect free. That means we can concentrate on the input and output behavior of the function under test for black-box testing (and the structure of the function, for glass-box testing). Regardless of the type of system we construct, there are some generally applicable rules that ease the testing of functional programs. For example, the rules from equivalence class testing [BIN 99] can be applied. Basically that means by looking at the function as a black-box and an informal specification one can identify input values that are treated uniformly by the function and make these an equivalence class. From each equivalence class it is regarded as sufficient to select one value and test this as a representative for the respective class. Furthermore, corner cases or extreme values like empty lists, empty strings, number zero and so on should be tested as well. It's also advisable to keep functions simple and avoid embedded lambda abstractions because anonymous functions are not (easily) testable. Functional programs usually are defined using rules with pattern matching. A coverage of these rules as well as their input patterns is advisable as well.

As we do not in generally deal with testing of functional programs, but with a special kind of programs generated from our state machine-based models, we now concentrate on the state machine-based testing approach, although we will discover that the principles discussed so far lead to very similar test cases. As said earlier, well-known test strategies for state machines do exist and can be reused.

As we described in section 8.3.3 our system includes a "runtime" part. Here we provide standard functions for example to execute state machines, to add a notion of time and to handle timers as addendum to state machines. Since this functionality remains the same regardless of the individual system and since we generate code against this functionality in a systematic way, we are able to ignore the runtime system for test case generation and concentrate on the specific part which resembles mainly the untimed state transition (delta) functions for the individual components. This leads to a more comprehensible and optimized set of test cases for each component. However, we also keep in mind that we also need to make system tests that check the functionality of the overall `abp` component.

For each component's delta function the test cases should fulfil transition coverage. That means every transition is executed at least once. Since we allow pre- and post-conditions in transitions, every expression in a disjunction is evaluated to True at least once. Note that this kind of decision coverage is not relevant for the ABP example as there are no disjunctions. In general disjunctions in these conditions should (and often can) be transformed by splitting the transition and handling the resulting transitions separately. If every state is reachable, transition coverage on state machines subsumes state coverage. However, we have to distinguish the finitely many states of the graphically depicted state machine and the potentially infinite number of states (and transitions) of the implementation. The relationship between both is handled by grouping the state space into equivalence classes using invariants. However, transition coverage now does not imply coverage of these equivalence classes anymore. Thus coverage criteria for state equivalence classes and transitions can be combined.

Beyond transition coverage, it is interesting to check the combined behavior of transitions, e.g. using full paths through the state machine. A minimized path coverage might check every path, where loops are only handled once (similar to "boundary-interior path tests" [NTA 88]). This technique is costly, as paths may be many, but also helpful, because some errors only occur in the combination of unusual paths, where nobody has been thinking of.

As a last issue to be considered, the input itself may be analyzed to derive possible tests. As we deal with streams of incoming messages, equivalence classes of messages may be considered, but also sequences of messages (e.g. what happens if the same message arrives twice?) or certain interleavings. This may lead to further refinements of the test cases. However, as variants of messages are usually handled through different pattern of inputs, coverage on input messages may be implied by coverage on transitions.

### 8.5. Implementing Tests in Haskell

Having clarified general considerations, we now show how to define test cases in a systematic way and show how they can be executed in an efficient manner, to allow us regression testing.

*8.5.1 Test Infrastructure*

Only a few generic functions are needed that serve as the test infrastructure similar to the runtime system discussed above. These functions later allow us to write test cases in a concise way, quite similar to unit test frameworks for other languages (e.g. [BEC 99b]). At first, we introduce a transition tester. `transT` takes the input and current state, executes a transition and compares the result with the expected state and output.

```
transT:: DeltaFct s i o -> (s, i , s,[o]) -> Bool
transT delta (s,i,expS,expO) = ((delta s i) == (expS, expO))
```

   Second, we introduce a path tester. The full path tester `pathT` checks whether a sequence of inputs leads to a certain path of states and sequence of outputs. Slightly adapted versions of a path tester just check the state or just the output. It's up to the test engineer to decide how fine granular a test needs to be defined. All versions take a transition function, a start state and a sequence of inputs. The implementation is straight-forward and omitted here since we're not going to give concrete examples for path tests in this paper. Nondeterministic transition functions are realized through an oracle, which allows us to fully control nondeterminism, but forces us to cover different oracles as well (not shown in the signatures below).

```
pathT  :: DeltaFct s i o -> (s, [i], [s,[o]]) -> [Bool]
pathTs :: DeltaFct s i o -> (s, [i], [s]) -> Bool
pathTo :: DeltaFct s i o -> (s, [i], [o]) -> Bool
```

*8.5.2. Tests for the ABP Components*

We will now illustrate the implementation of test cases in Haskell considering the sender component as example. Our goal is to manually derive transition coverage for the sender and show how these transitions can be denoted easily. The sender component has a total of eight transitions that need to be tested. This leads to eight tests to cover the state machine transitions.

We start by explaining how a test case that covers a single transition can be derived from the state machine specification. As an example, consider the following transition from figure 8.2:

```
{b2 = b, #s ≥ 2}
 b2 / (b, head (tail s)); SetTimer 3
{b' = ¬b, s' = tail s}
```

Furthermore, the transition's source and destination state is characterized by a non-empty buffer. First, we need to identify a valid start state for the transition. Since the buffer needs to be non-empty and, due to the pre-condition, at least two messages long, one possible start state can be `(True, [3,4])`. The transition's input is a bit `b2` whose value is restricted by the pre-condition. To test the specification, we derive the appropriate resulting state and output not from the model, but determine from our background knowledge what must happen. In this setting it is necessary to act as test oracle ourselves, as we are going to check the specification. We merely analyze the transition system to understand, what transitions need to be covered. In this case, we derive the resulting state `(False, [4])` and expected output `[(False,4),SetTimer 3]`. The rest of the sender transitions can be handled in the same manner. A further detailing is not necessary, because neither an oracle, nor internal complicated pre/postconditions occur, nor is the sender incompletely specified. We collect all transition tests in a table as follows.

| no. | source state | input | destination state | output |
|---|---|---|---|---|
| 1 | (True,[]) | True | (True,[]) | [] |
| 2 | (True,[]) | 3 | (True,[3]) | [(True,3),SetTimer 3] |
| 3 | (True,[3]) | 4 | (True,[3,4]) | [] |
| 4 | (True,[3,4]) | True | (False,[4]) | [(False,3),SetTimer 3] |
| 5 | (True,[4]) | True | (False,[]) | [SetTimer (-1)] |
| 6 | (True,[3,4]) | False | (True,[3,4]) | [] |
| 7 | (True,[3,4]) | TimeoutEvent | (True,[3,4]) | [(True,3),SetTimer 3] |
| 8 | (True,[]) | TimeoutEvent | (True,[]) | [] |

Please note that like in the defining state transition diagram, we do not need to tag the incoming input, because the values on both channels are disjoint. Boolean values are acknowledgements and integers are used as messages. As discussed, the corresponding Haskell definitions need to deal with this union of channels, resulting in a more awkward and less readable definition. Let us assume they are given in a list called `senderTransitionTests`. The defined transition tests can systemati-cally be mapped to Haskell. Together with the earlier mentioned functions (and possible path tests for the sender) a test suite is defined easily:

```
senderTestSuite = map (transT senderDelta) senderTransitionTests
```

To execute the sender test we consequently only need to evaluate `senderTestSuite`, receiving a larger list of Booleans (hopefully all `True`. Using the definition

```
all = and (senderTestSuite ++ receiverTestSuite ++ ...)
```

allows us to resemble the well known "green/red"-light from unit testing. Due to space limitations, only the test case definition and execution for the sender component is shown. Analogously, component tests for the medium and receiver can be derived. However, the media exhibit special characteristics that we have not dealt with so far, as they use an oracle. As we can understand the oracle as a special case of input sequence, we therefore just need to analyze possible interesting inputs sequences and run those together with the other tests.

We showed how to systematically define and execute test cases for delta functions of state machines. Especially for system level testing of the composed ABP function it might also be useful to generate a larger set of test cases randomly. Since this is not in the focus of this paper we refer the reader for example to [CLA 00, KOO 03].

### 8.6. Discussion of Results

In this paper, we have discussed an approach to model behavior of distributed asynchronously communication behavior. To test these models we had to map them into an executable form. We chose the functional language Haskell for that purpose, because Haskell offers lazy lists as well as pattern matching techniques, which perfectly allow us to simulate our underlying semantics.

As a next step we have understood, how tests cover transitions, state, input or even paths. This allows us to systematically derive tests. However, our approach of an automatic mapping ("code generation") of the model into the simulation engine does not allow us to derive complete tests from the model. Deriving code and tests from the same model does not allow us to check correctness of the model, but consistency of the generators. Thus our approach so far only allows us to understand, what test inputs are of specific interest, but forces us to manually add the desired test result to the test. The situation changes, when we do a manual implementation of the model. Then this simulation engine can be used to derive test results that can be used as test oracles for the actual implementation.

When applying this approach to other distributed systems, we found the approach very effective for us developers. Through systematic and early definition of tests, we found some subtle errors very early in the specification model already. Deriving tests from requirement and design models is worth the effort – particularly in complex distributed settings and when validating protocols.